\documentclass[prl,twocolumn,showpacs,amsmath,amssymb,superscriptaddress]{revtex4}
\usepackage{graphicx}
\usepackage{dcolumn}
\usepackage{bm}
\usepackage{natbib}
\usepackage{amssymb}
\usepackage{dcolumn}
\usepackage{bm}
\usepackage[latin2]{inputenc}
\usepackage{epsfig}
\usepackage{amsmath}
\usepackage{amssymb}
\usepackage{dsfont}
\usepackage{dcolumn}
\usepackage{natbib}
\usepackage{dcolumn}

\newcommand{\kk}{{\bf k}}

\begin{document}

\title{Systematic study of Mn-doping trends in optical properties of (Ga,Mn)As}

\author{T. Jungwirth}
\affiliation{Institute of Physics ASCR, v.v.i., Cukrovarnick\'a 10, 162 53 Praha 6, Czech Republic}
\affiliation{School of Physics and Astronomy, University of Nottingham, Nottingham NG7 2RD, United Kingdom}

\author{P.~Horodysk\'a}
\author{N.~Tesa\v{r}ov\'a}
\author{P.~N\v{e}mec}
\author{J.~\v{S}ubrt}
\author{P.~Mal\'y}
\affiliation{Faculty of Mathematics and Physics, Charles University in Prague, Ke Karlovu 3,
121 16 Prague 2, Czech Republic
}
\author{P. Ku\v{z}el}
\author{C. Kadlec}
\author{J. Ma\v{s}ek}
\affiliation{Institute of Physics ASCR, v.v.i., Na Slovance 2, 182 21 Prague 8, Czech Republic}

\author{I.~N\v{e}mec}
\affiliation{Faculty of Science, Charles University in Prague, Hlavova 2030, 128 40 Prague 2, Czech Republic}

\author{V. Nov\'ak}
\affiliation{Institute of Physics ASCR, v.v.i., Cukrovarnick\'a 10, 162 53 Praha 6, Czech Republic}
\author{K.~Olejn\'{\i}k}
\affiliation{Institute of Physics ASCR, v.v.i., Cukrovarnick\'a 10, 162 53 Praha 6, Czech Republic}
\affiliation{Hitachi Cambridge Laboratory, Cambridge CB3 0HE, United Kingdom}
\author{Z.~\v{S}ob\'a\v{n}}
\affiliation{Institute of Physics ASCR, v.v.i., Cukrovarnick\'a 10, 162 53 Praha 6, Czech Republic}
\affiliation{Faculty of Electrical Engineering, Czech Technical University in Prague, Technick\'a 2, 166 27 Prague, Czech Republic}
\author{P.~Va\v{s}ek}
\author{P.~Svoboda}
\affiliation{Institute of Physics ASCR, v.v.i., Cukrovarnick\'a 10, 162 53 Praha 6, Czech Republic}

\author{Jairo Sinova}
\affiliation{Department of Physics, Texas A\&M University, College Station, Texas 77843-4242, USA}
\affiliation{Institute of Physics ASCR, v.v.i., Cukrovarnick\'a 10, 162 53 Praha 6, Czech Republic}

\begin{abstract}
We report on a systematic study of optical properties  of (Ga,Mn)As  epilayers  spanning the wide range of accessible substitutional Mn$_{\rm Ga}$ dopings. The growth and post-growth annealing procedures  were optimized for each nominal Mn doping in order to obtain films which are as close as possible to uniform uncompensated (Ga,Mn)As mixed crystals. We observe a broad maximum  in the mid-infrared absorption spectra whose position exhibits a prevailing blue-shift for increasing Mn-doping.  In the visible range, a peak in the magnetic circular dichroism  blue shifts with increasing Mn-doping. These observed trends confirm that disorder-broadened valence band states provide a better one-particle representation for the electronic structure of high-doped (Ga,Mn)As with metallic conduction than an energy spectrum assuming the Fermi level pinned in a narrow  impurity band. 
\end{abstract}

\date{\today}
\maketitle

The discovery of ferromagnetism in (Ga,Mn)As above 100~K  \cite{Ohno:1998_a} opened an attractive prospect for exploring the physics of magnetic phenomena in doped semiconductors and for developing advanced concepts for spintronics. Assessment  of a wide range of magnetic and transport properties of the material \cite{Matsukura:2002_a,Jungwirth:2006_a,Dietl:2008_b} showed that in  ferromagnetic (Ga,Mn)As with Mn dopings $x>1$\%, disorder-broadened and shifted host Bloch bands  represent a useful one-particle basis for describing this mixed-crystal degenerate semiconductor. The common kinetic-exchange model implementation of this valence band theory and the more microscopic tight-binding Anderson model or {\em ab-initio} density functional theory can all be shown \cite{theory_paper} to be  mutually consistent  on the level of atomic and orbital resolved  band structure.  The main utility of valence band theories have been in providing a qualitative and often semi-quantitative description of phenomena originating from the exchange split and spin-orbit coupled electronic structure and in assisting the development of prototype spintronic devices \cite{Dietl:2008_b}. Other basic physical properties of (Ga,Mn)As, namely those reflecting the vicinity of the metal-insulator transition and localization and electron-electron interaction effects, remain to be fully understood and require to go beyond the commonly employed perturbative or disorder averaged  Bloch-band theories.

In the insulator non-magnetic regime ($x\ll1$\%), the system is readily described by localized Fermi level states residing inside a narrow impurity band separated from the valence band by an energy gap of magnitude close to the isolated Mn$_{\rm Ga}$ impurity binding energy. Recently, a debate has been stirred by proposals, based in particular on optical spectroscopy measurements  \cite{Burch:2008_a}, that the narrow impurity band persists in high-doped (Ga,Mn)As with metallic conduction.  
Several phenomenological variants of the impurity band model have been proposed for the high-doped  regime \cite{Burch:2006_a,Stone:2008_a,Ando:2008_a,Tang:2008_a,Burch:2008_a} which are mutually inconsistent from the perspective of the assumed atomic orbital nature of the impurity band states~\cite{theory_paper}. Further theoretical inconsistencies arise when recreating the phenomenological models microscopically with the constraint of the experimentally determined moderate binding energy of an isolated Mn$_{\rm Ga}$  of $E_a=0.1$~meV. A detached narrow impurity band does not persist in any of the variants of the model to dopings $x>1$\%  corresponding to (Ga,Mn)As with metallic conduction \cite{theory_paper}.

The goal of this paper is to resolve the controversy between valence and impurity band approaches to  (Ga,Mn)As from the experimental perspective by assessing doping trends in the optical spectra over a wide Mn concentration range in a consistently and controllably  prepared set of materials. We have optimized the growth and post-growth annealing procedures  individually for each nominal doping in order  to  minimize the density of compensating defects and other unintentional impurities and to achieve high uniformity of the epilayers in the growth and lateral directions, as detailed in the Supplementary material  \cite{suppl}. The trends observed in the unpolarized infrared and visible-range magneto-optical spectroscopies of this systematically prepared set of samples support the disordered valence band approach. They contradict previous conclusions which have been based on experiments  in a more limited set of (as-grown and annealed ) samples  and which have been regarded as the key evidence for impurity band conduction in metallic (Ga,Mn)As \cite{Burch:2006_a,Ando:2008_a,Tang:2008_a, Burch:2008_a}.   

\begin{figure}[ht]
\vspace*{-0.2cm}
\hspace{0cm}\includegraphics[width=1\columnwidth,angle=0]{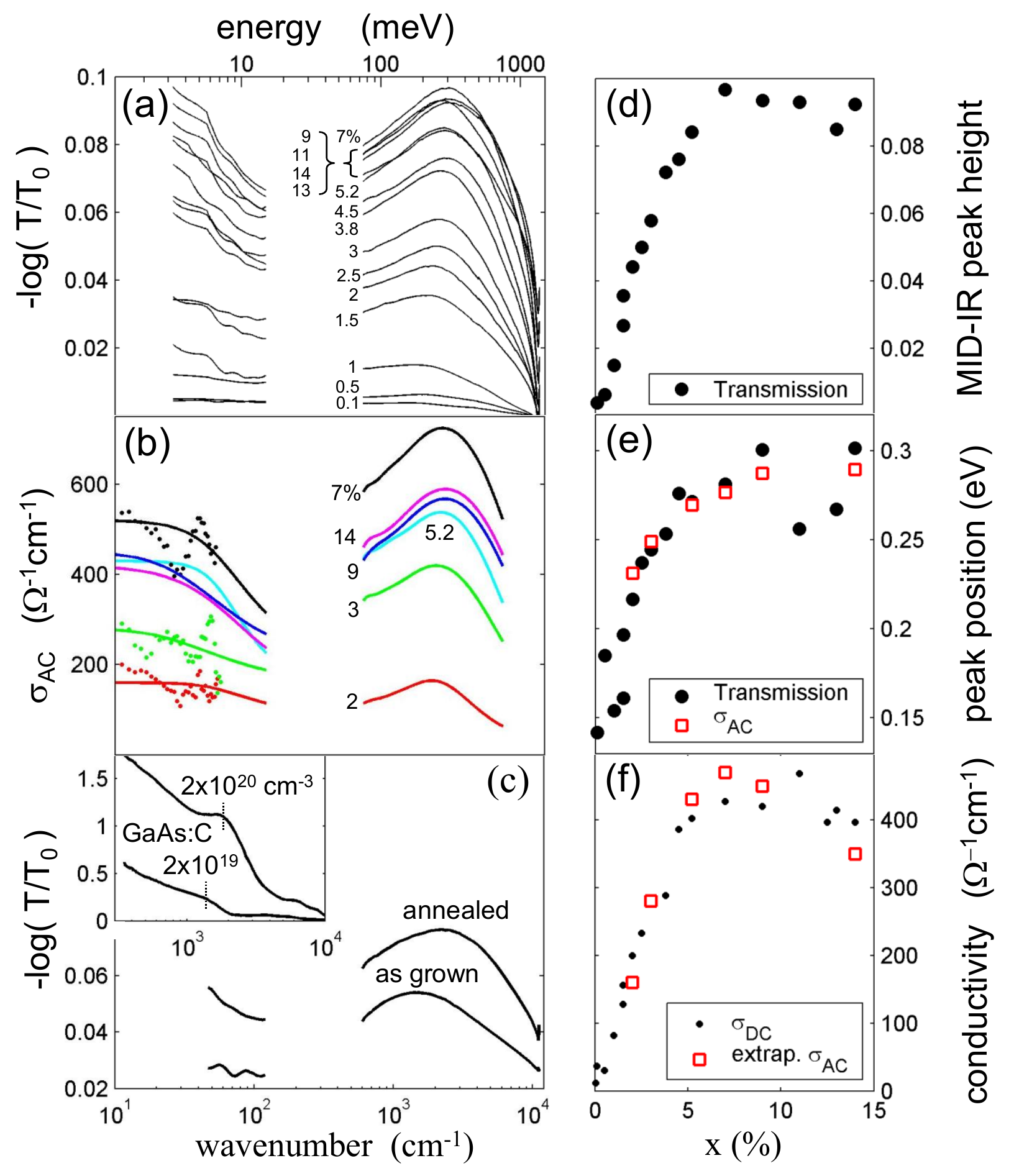}
\vspace*{-0.5cm}
\caption{(a) Infrared absorption of a series of optimized (Ga,Mn)As/GaAs epilayers with nominal Mn doping $x=0.1-14$\% plotted from the measured optical transmissions of the samples ($T$) and of the reference bare GaAs substrate ($T_0$). Spectra of the 100~nm thick samples with $x\le 1$\% were divided by 5 for the consistency with those measured for the 20~nm thick higher doped samples. (b) Real part of the ac conductivity (lines) obtained from the measured complex conductivity in the terahertz range (points) and from fitting the complex conductivity in the infrared range to the measured transmissions. (c) Comparison of the infrared absorption in as-grown and annealed 4.5\% doped sample. Inset: Comparison to GaAs:C samples with carbon doping densities $2\times 10^{19}$~cm$^{-3}$ and $2\times 10^{20}$~cm$^{-3}$. (d) Height of the (Ga,Mn)As mid-infrared absorption peak as a function of Mn doping. (e) Position of the peak inferred from the transmission measurements and from the fitted ac conductivities. (f) Zero frequency conductivities obtained from dc transport measurements and from extrapolated optical ac conductivities measured in the terahertz range.}
\label{fig1}
\end{figure}

Nominal dopings of our set of (Ga,Mn)As epilayers grown on GaAs substrates span a range from paramagnetic insulating materials  with $x<1\%$ to  materials with $x$ up to $\approx 14$\%, corresponding to $\approx 8$\% of uncompensated Mn$_{\rm Ga}$, and ferromagnetic transition temperatures reaching 190~K \cite{suppl}. Samples with $x\le1\%$ have thickness of 100~nm. All epilayers with $x\ge1.5\%$ are 20~nm thick. In these materials with large Mn doping, high quality epilayers are obtained only for thicknesses larger than $\sim10$~nm and lower than $\sim50$~nm. All samples within the series have  reproducible characteristics with the overall trend of  increasing Curie temperature (in the ferromagnetic films), increasing hole concentration, and increasing magnetic moment density with increasing $x$. The samples have a high degree of uniformity on a macroscopic scale as inferred from their  sharp magnetic and transport singularities at the Curie point \cite{suppl}. 

Samples used in transmission measurements have a polished back side of the wafers to minimize diffusive light scattering. The unpolarized transmission experiments \cite{suppl} on the (Ga,Mn)As/GaAs samples and on the control bare GaAs substrate were performed at 300~K in the range 25--11000~cm$^{-1}$ (3--1360~meV) using the Fourier transform infrared spectroscopy. The range between 120-600~cm$^{-1}$ (15--75~meV) with strong phonon response in GaAs substrate is excluded from the data. Measurements of the complex conductivity in the  low-frequency range  8--80~cm$^{-1}$ (1--10~meV) of the (Ga,Mn)As/GaAs wafers (with measurement of the bare GaAs substrate as a reference) were performed by means of terahertz time-domain transmission spectroscopy \cite{suppl}. Magneto-optical experiments \cite{suppl} at 15~K in the near infrared to visible range 970--20160~cm$^{-1}$ (1.2--2.5~eV) were performed primarily in the reflection geometry because of the small epilayer thickness of the optimized materials. Control magneto-optical measurements in transmission were done on a 230~nm thick (Ga,Mn)As epilayer with the GaAs substrate removed after growth by wet etching.

First we discuss the observed broad  maximum in the unpolarized infrared absorption near 200~meV. In Fig.~1(a) we plot experimental data obtained directly from the measured infrared transmissivities. Real part of the ac conductivity curves shown in Fig.~1(b) represent the best fit to the measured transmission  in the THz and infrared ranges \cite{suppl}. The fit is anchored at low-frequencies by the directly measured THz conductivity. (The scatter in the measured THz conductivity reflects the precision of these measurements which is limited primarily by the quality of sample surfaces \cite{suppl}.) The position of the mid-infrared absorption peak in both representations of the measured data has a prevailing  blue-shift tendency with increasing doping. This is reminiscent of the blue-shift of this spectral feature seen in our (and previously studied \cite{Songprakob:2002_a}) control  GaAs:C materials, shown in Fig.~1(c). We recall that  for the non-magnetic hydrogenic acceptors it is established that the peak originates from transitions inside the semiconductor valence band. Based on microscopic valence band theory calculations,  these transitions have been also predicted to yield the broad maximum in the mid-infrared spectra of (Ga,Mn)As  which is rather insensitive to the exchange splitting of the valence band \cite{Sinova:2002_a,Yang:2003_b}. Apart from the predicted prevailing blue shift, the  peak position in (Ga,Mn)As can have a non-monotonic dependence on doping as a consequence of momentum non-conserving transitions allowed by the strong disorder in (Ga,Mn)As~\cite{Jungwirth:2007_a}. 

A red shift of the measured mid-infrared peak reported in Ref.~\cite{Burch:2006_a} in as-grown and annealed samples arranged by expected increasing hole concentrations has been presented as the key evidence of the failure of the valence band theories. The data were phenomenologically interpreted in terms of  impurity band conduction persisting to the high-doped metallic (Ga,Mn)As. In the subsequent analysis it has been pointed out \cite{Jungwirth:2007_a}, however, that the association of this peak to an impurity band is implausible (i) because of the absence of the thermally activated dc-transport counterpart in the high-doped samples, (ii) because of the initial blue-shift of this  mid-infrared feature with respect to the impurity band transition peak in the very dilute insulating samples, and (iii) because of the appearance of the peak at frequencies above $2E_a$ which is the expected upper bound for impurity band transitions. Our data in Fig.~1 corroborate the conclusions of Ref.~\cite{Jungwirth:2007_a} by demonstrating that the red-shift of the mid-infrared peak with increasing Mn$_{\rm Ga}$ doping is not the general trend  in  (Ga,Mn)As materials prepared with the minimized number of compensating and other unintentional impurities. 

In  Figs.~1(a) and (b) we have not included measurements in the as-grown epilayers  because their characteristics are not fully reproducible, are more ambiguous, and the materials are less uniform \cite{suppl}. Nevertheless, to make a connection to previous studies we have measured the mid-infrared peak in a nominally 4.5\% Mn-doped sample before and after annealing. As shown in Fig.~1(c), we observe a blue shift of the mid infrared peak after annealing, i.e., the red-shift is not observed in our materials even if we use annealing  to increase the effective doping. The same trend was confirmed in our as-grown and annealed  samples with 12\% nominal Mn doping. 

In Fig.~1(d)-(f) we highlight the observed correlation between the peak position and its height, as well as the expected correlation from the valence band theories between the amplitude of the peak and the dc conductivity. Note that in (Ga,Mn)As, the dc conductivity initially increases with doping  because of the increasing number of itinerant holes provided by the Mn acceptors. At  large dopings, the competition of increasing hole concentration and decreasing mobility yields a non-monotonic dc and ac conductivity trend (see also Ref.~\cite{suppl}). We also point out the close correspondence shown in Fig.~1(f) between the dc conductivities obtained from extrapolated THz data and from dc longitudinal and Hall transport measurements, which confirms the consistency of our optical data.

\begin{figure}[ht]
\includegraphics[width=1\columnwidth,angle=0]{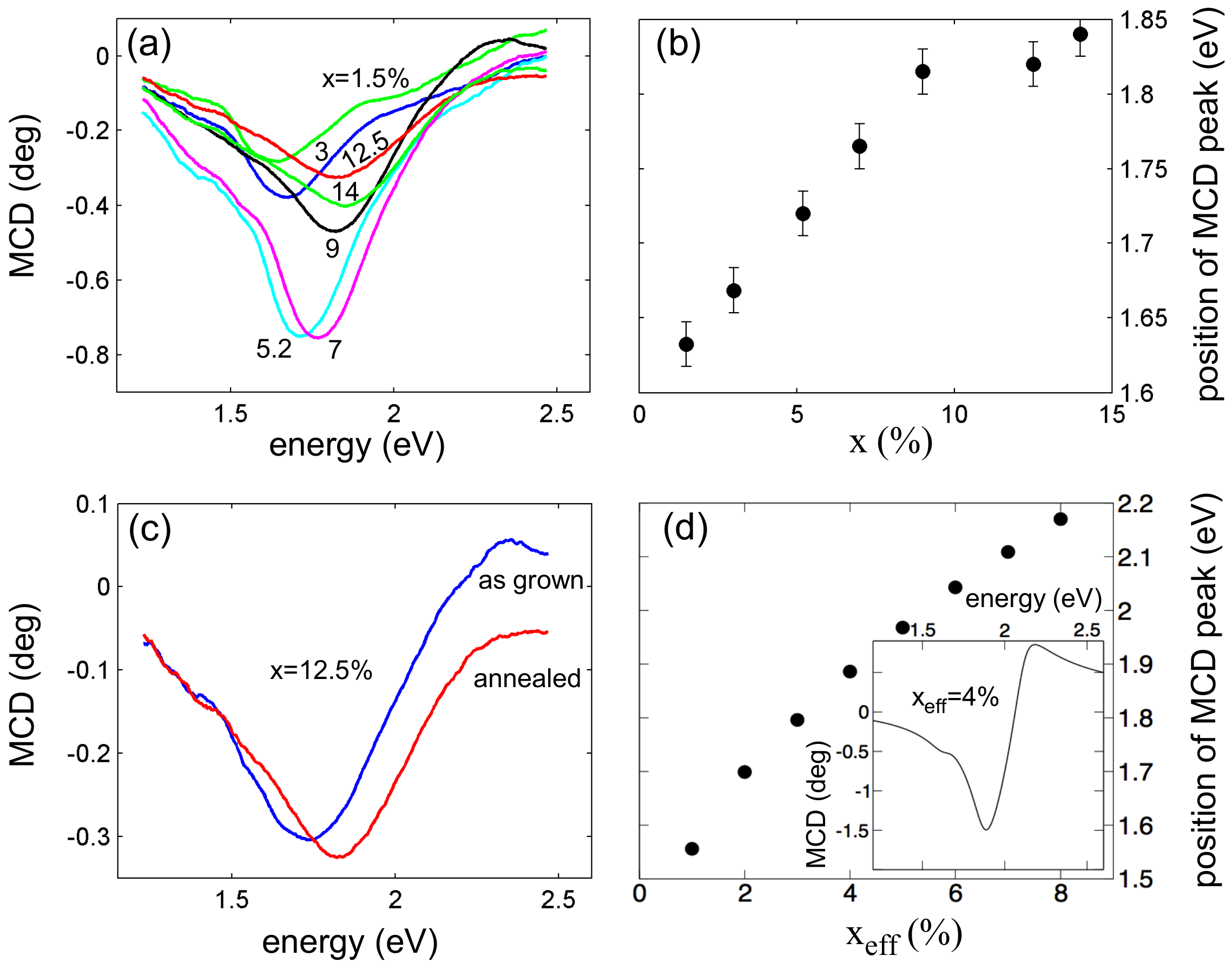}
\vspace*{-0.5cm}
\caption{(a) Reflection MCD measurements in samples from the optimized series of 20~nm thick epilayers spanning the whole studied range of Mn dopings of ferromagnetic  (Ga,Mn)As. (b) Experimental position of the MCD peak (negative in the considered sign convention) as a function of nominal doping. (c) Comparison of the MCD measurements in the 12.5\% doped as-grown and annealed epilayer. (d) Theoretical position of the MCD peak as a function of effective doping of uncompensated Mn$_{\rm Ga}$ impurities (see Ref.~\cite{suppl}); inset shows theoretical MCD spectrum  for 4\% effective doping.}
\label{fig2}
\end{figure}

\begin{figure}[ht]
\vspace*{-0.2cm}
\hspace*{-0cm}\includegraphics[width=1\columnwidth,angle=0]{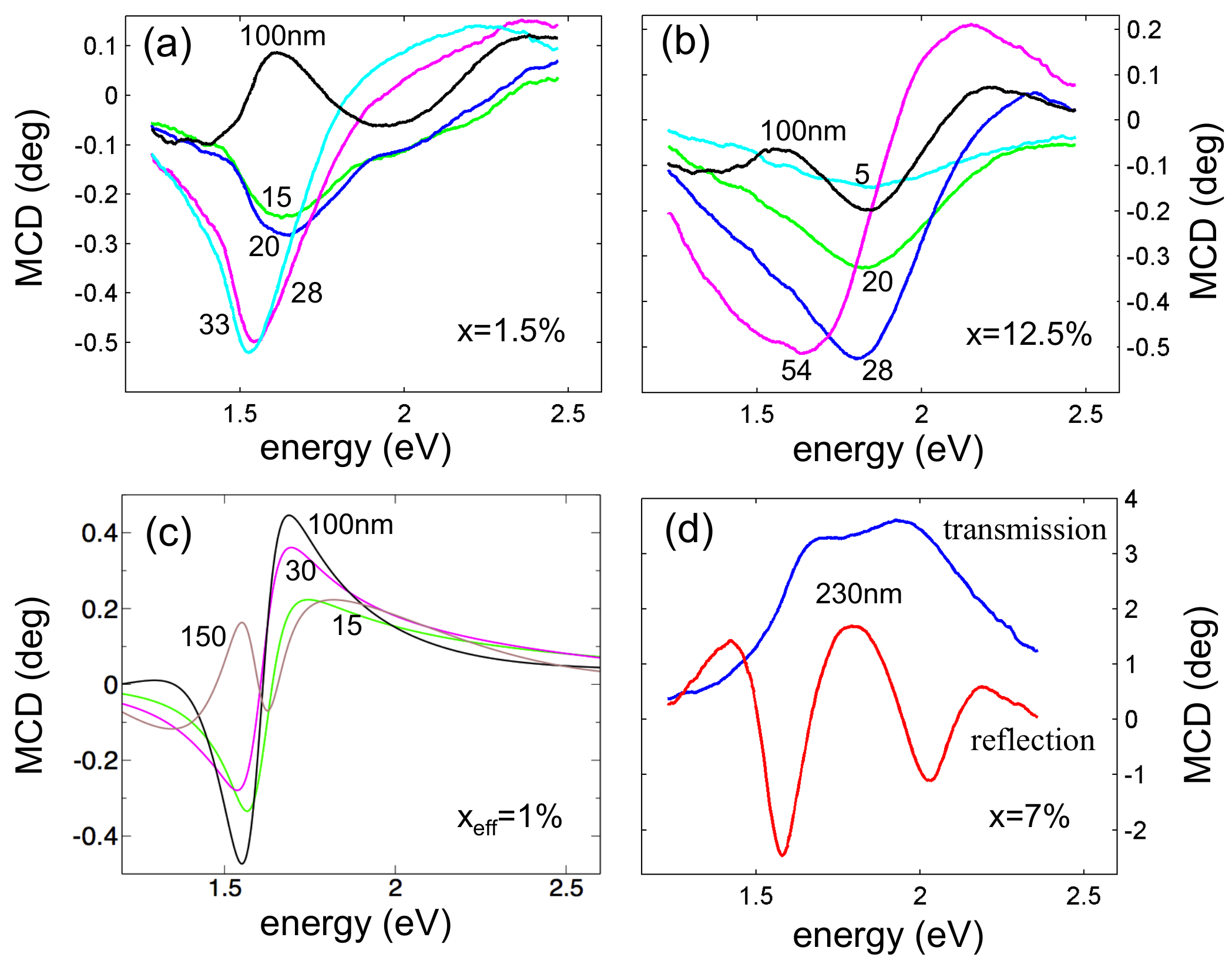}
\vspace*{-0.5cm}
\caption{ Measured reflection MCD data in samples with different (Ga,Mn)As epilayer thickness and nominal Mn doping 1.5\% (a), 12.5\% (b). (c) Theoretical thickness dependence of the MCD spectra at 1\% effective doping. (d) Transmission and reflection MCD measurements in a 7\% nominally doped 230~nm thick free standing (Ga,Mn)As epilayer.}
\label{fig3}
\end{figure}

Unlike the unpolarized optical spectroscopy, the magneto-optical effects are very sensitive to the magnetic state of the system, their interpretation is  more microscopically constrained, and they yield sharper spectral features. This is particularly valid for the infrared magneto-optical spectroscopy for which previous study \cite{Acbas:2009_a} has shown that the disordered valence band  theory with kinetic-exchange-split bands of ferromagnetic (Ga,Mn)As accounts semi-quantitatively for the overall characteristics of the measured data. We now extend the analysis to higher energies including transitions across the semiconductor band gap. 

The magnetic circular dichroism in the reflection geometry is given by ${\rm MCD(deg)}=90/\pi\,(R^+-R^-)/(R^++R^-)$, where $R^{+(-)}$ is the reflected intensity of the light with the angular momentum +1 (-1). The main result of MCD measurements  in our (Ga,Mn)As materials, shown in Figs.~2(a) and (b), is the observed  large blue-shift of the MCD peak (the peak is negative in the above sign convention). Our data contradict the conclusion based on experiments in Ref.~\cite{Ando:2008_a} which stated that the position of the MCD peak was independent of doping and that this was a signature of the pinning of the Fermi level in a rigid narrow impurity band  \cite{Ando:2008_a,Tang:2008_a}. As in the case of the unpolarized infrared optical measurements, we have performed additional MCD experiments in pairs of as-grown and annealed (Ga,Mn)As materials. Upon annealing, we again observe a blue-shift of the MCD peak, as illustrated in Fig.~2(c).

Theoretical modeling of the MCD spectra within the valence band theory, described in detail in Ref.~\cite{suppl}, is shown in Fig.~2(d). The magnitude of the peak and its absolute position depend on the detailed implementation of the model, e.g., on  the way how disorder or band-gap renormalization effects are treated \cite{suppl}. On a qualitative level and independent of these specific implementations of the model, the valence band theory with the antiferromagnetic $p-d$ kinetic exchange \cite{Jungwirth:2006_a}  reproduces the overall experimental shape of the MCD spectrum, the sign of the MCD peak, and the blue shift of the peak with increasing doping. 

Because our experiments were performed in the reflection geometry we have prepared several control samples to assess the role of multiple reflections.  In Figs.~3(a),(b) we show a comparison of MCD spectra in (Ga,Mn)As epilayers of different thicknesses controlled during growth or post-growth by etching \cite{suppl}. In the control thicker (Ga,Mn)As samples, the MCD peak is superimposed on oscillations caused by  the multiple reflections. The oscillations are completely suppressed  in the 20~nm thick films and the reflection MCD peak position is therefore not affected by the film thickness in our series of optimized 20~nm thick (Ga,Mn)As epilayers. 
The transition from a thin film to a film where multiple reflections are important is captured qualitatively by the valence band theory calculations, as shown in Fig.~3(c).

As an additional consistency check of the sign of our MCD signals with respect to previous MCD experiments in (Ga,Mn)As we have also measured the  free-standing 230~nm thick 7\% Mn-doped (Ga,Mn)As epilayer in both reflection and transmission MCD geometry. In the latter case the dichroism is given by ${\rm MCD(deg)}=90/\pi\,(T^+-T^-)/(T^++T^-)$, where $T^{+(-)}$ is the transmitted intensity of the light with the angular momentum +1 (-1). Our transmission MCD data shown in Fig.~3(d) are very similar to measurements in samples with comparable doping and film thickness reported in Ref.~\cite{Beschoten:1999_a}. 

To conclude, we have not observed the signatures in the optical spectra which in previous studies have been interpreted as general trends in metallic (Ga,Mn)As and as an evidence of  Fermi level states residing in a narrow impurity band detached from the valence band. Our unpolarized infrared and visible-range magneto-optical measurements performed on an extensive series of systematically prepared (Ga,Mn)As materials support the valence band theory representation of the effective one-particle spectrum of high-doped  (Ga,Mn)As with metallic conduction. 

\acknowledgments
We acknowledge support  from EU Grants FP7-215368 SemiSpinNet and FP7-214499 NAMASTE, from Czech Republic Grants  KAN400100652, LC510,  MEB020928, Preamium Academiae, from the research plans AV0Z10100521, MSM0021620834, MSM0021620857, Grant Agency of the Czech Republic Grant 202/09/H041, Grant SVV-2010-261306 of the Charles University in Prague, and from U.S. Grants  ONR-N000140610122,  DMR-0547875, and SWAN-NRI. JS is a Cottrell Scholar of Research Corporation.


\newpage
\onecolumngrid
\begin{center}
{\Large\bf SUPPLEMENTARY MATERIAL}
\end{center}
\vspace*{3.3cm}  
\section{Introduction}
The supplementary material describes a series of (Ga,Mn)As epilayers intended to provide a basis for systematic experimental studies of this ferromagnetic semiconductor. For each nominal Mn doping $x$, the growth and post-growth annealing conditions were separately optimized in order to achieve the highest Curie temperature $T_c$ attainable at the particular $x$. The highest $T_c$ criterion was found to lead simultaneously to layers with maximized uniformity and  minimized compensation by unintentional impurities and defects.

The nominal doping $x$ spans a wide range from non-ferromagnetic insulating materials  with $x<1\%$ to the highest $T_c$ materials with $x\approx 13\%$. Within this nominal Mn doping range and for film thicknesses smaller than approximately 50~nm and larger than 10~nm in case of  higher doped samples with $x\ge1.5$, we obtained a series of (Ga,Mn)As materials with reproducible characteristics. They show an overall trend of increasing saturation moment (ascribed to substitutional Mn$_{\rm Ga}$) with increasing $x$ , increasing $T_c$, and increasing hole density. The materials have no apparent charge or moment compensation  of the substitutional Mn$_{\rm Ga}$ impurities and have a large degree of uniformity reflected by sharp magnetic and transport singularities at $T_c$. 

Because experimentally it is difficult to determine on a precise quantitative level  all relevant parameters of a given material and also because accurate quantitative theoretical modeling of even an idealized (Ga,Mn)As is not feasible we are convinced that a meaningful examination of intrinsic properties of (Ga,Mn)As ferromagnetic semiconductor materials should be performed on the level of qualitative or at most semi-quantitative trends observed over a wide doping range in a consistent set of materials. Our series of epilayers  represents  an attempt to provide such an experimental basis. The optimization of the materials in the series, which is performed individually for each nominal doping, minimizes the uncertainties in the experimental sample parameters and produces high quality epilayers which are as close as possible to to uniform uncompensated (Ga,Mn)As mixed crystals. We prefer this method for studying the doping trends over, e.g., comparisons of as-grown and annealed samples because the as-grown state of a sample is not well defined; the characteristics of as-grown samples are not well reproducible, the materials are less uniform and can have a large number of unintentional interstitial Mn impurities which can compensate both the local moment and holes produced by   Mn$_{\rm Ga}$ and whose density is difficult to measure accurately. Similarly, the approach based on individual optimization at each nominal doping provides a qualitatively better control over the whole set of sample parameters and therefore a better  approximation to idealized (Ga,Mn)As materials than the combinatorial method which on one physical wafer produces a wide range of samples due to the  stoichiometry gradients across the wafer imposed during a single growth experiment. For further discussion highlighting the importance of the individual optimization of material synthesis at each nominal doping see also Ref.~\cite{Wang:2008_e}.

\section{Growth}
The growths were done in a Veeco Gen II molecular-beam epitaxy (MBE) system. We used 2-inch semi-insulating 0.5~mm thick GaAs substrates mounted on a molybdenum In-free sample holder by mechanically fixing the sample at the edges. Each growth started by a standard high-temperature grown GaAs buffer of about 200 nm thickness. After the growth interruption the substrate was cooled down for the low-temperature growth of (Ga,Mn)As; during the cooling period the As source was closed when the substrate temperature dropped below 400$^\circ$C.

The (Ga,Mn)As layers were grown at the growth rate of 0.2 monolayers/second, corresponding to temperatures of 940$^\circ$ and
920$^\circ$C at the tip and the base zone, respectively, of the dual filament Ga source. The Mn flux, and thence $x$, was determined by measuring the ratio of the beam equivalent pressures (BEP) of Mn and Ga sources before each growth. The Mn content was cross-checked on several samples by secondary ion mass spectroscopy (SIMS) and by comparing the growth rates of GaAs and (Ga,Mn)As; the results of the three methods agreed within $\Delta x\sim 0.015$. The BEP-based method yields the best reproducibility and guarantees the correct ordering by increasing $x$ even though the relative uncertainty in $x$ may be large especially at low dopings.

The Arsenic was supplied by a valved cracker cell in the As$_4$ regime (cracking zone temperature 600$^{\circ}$C). The As-to-(Ga+Mn) flux ratio was calibrated using the As-controlled RHEED intensity oscillations. For this purpose the substrate temperature was first set to 500$^{\circ}$C. Next, the As source was closed and the Ga shutter was simultaneously opened for 30 -- 40~s allowing accumulation of several Ga monolayers on the substrate surface. After closing the Ga shutter and re-opening the As source several growth oscillations were observed in the RHEED image with period corresponding to the As flux. We have checked that this technique yields essentially the same calibration as the As-rich to Ga-rich surface reconstruction transition method used in Ref.~\cite{Wang:2008_e}.

\begin{figure}[!h]
\vspace*{0cm}
\hspace*{-0cm}\includegraphics[width=.8\columnwidth,angle=0]{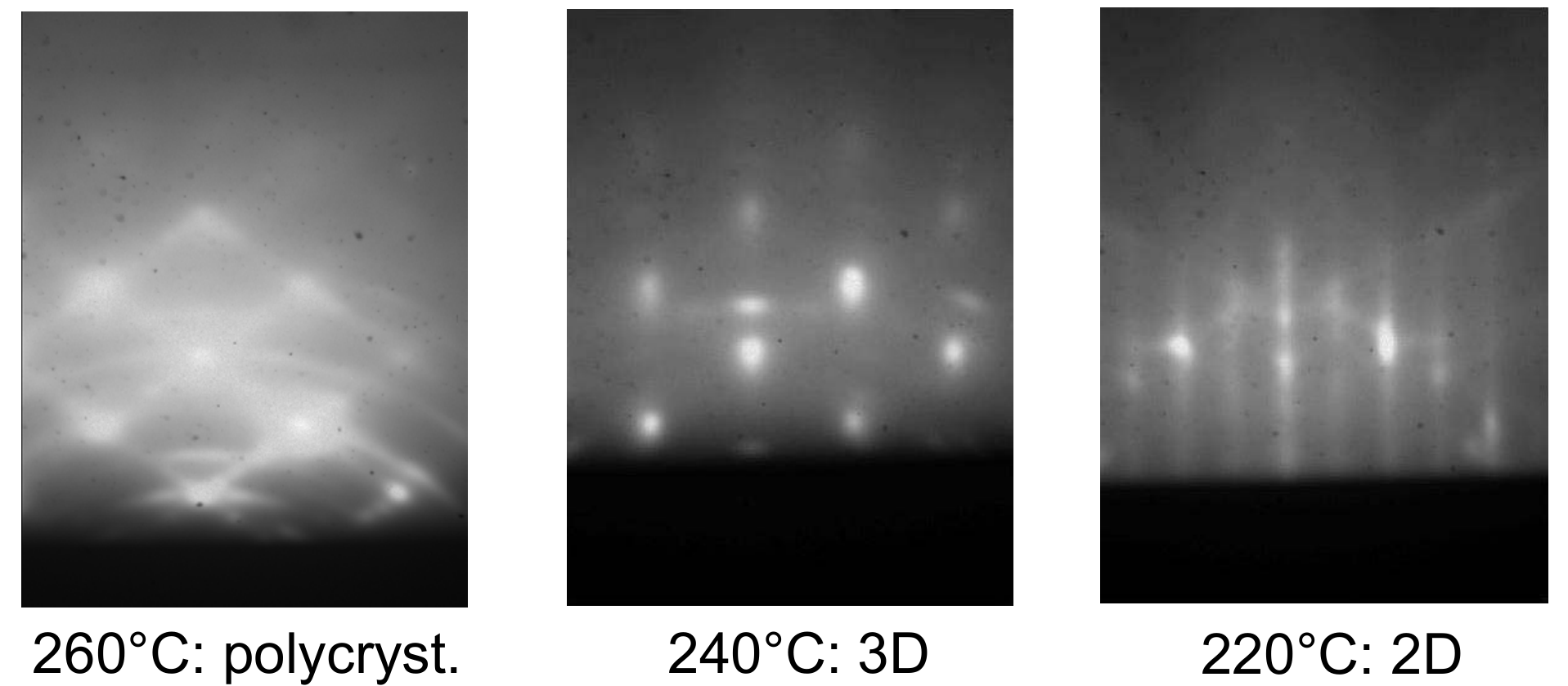}
\vspace*{0cm} \caption{RHEED images taken during growths of (Ga,Mn)As materials with nominal 7\% Mn doping, stoichiometric flux ratio and different substrate temperatures $T_s$. The 2D growth mode is achieved only at sufficiently low $T_s$.}
\label{fig_rheed}
\end{figure}

\begin{figure}[!h]
\vspace*{0cm}
\hspace*{-0cm}\includegraphics[width=0.4\columnwidth,angle=0]{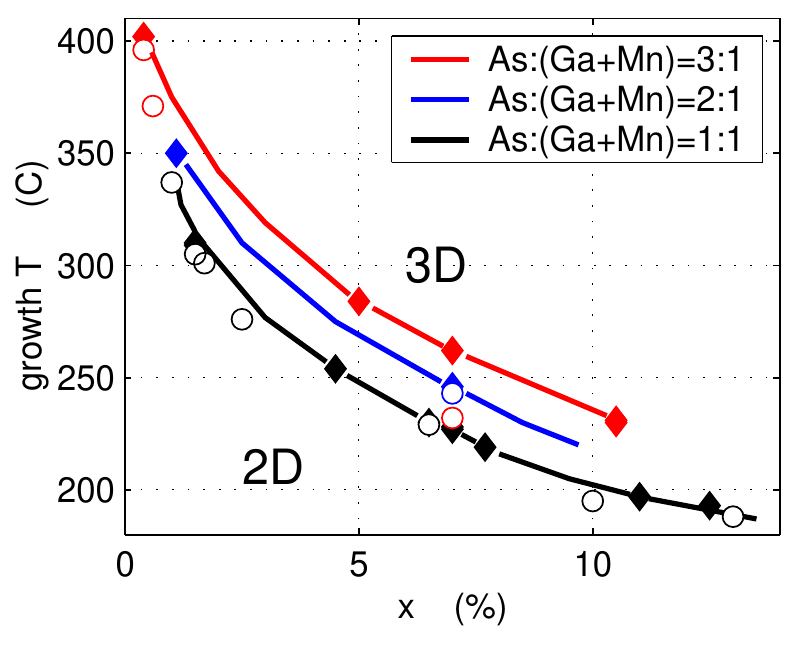}
\vspace*{0cm} \caption{Boundary in substrate growth temperature separating the 3D and 2D growth modes as a function of the nominal doping. Bottom (black), middle (blue) and upper (red) curves correspond to As:(Ga+Mn) ratio of 1:1, 2:1, and 3:1, respectively. Empty circles denote samples which retained the 2D growth mode at thickness 50~nm and more, full diamonds denote samples which became 3D at thicknesses between 20 and 50~nm.}
\label{fig_3D-2D}
\end{figure}

An extremely critical growth parameter of (Ga,Mn)As is the substrate temperature. For its measurement we used the GaAs band edge spectrometer (kSA BandiT) mounted on the central pyrometer port, normal to the substrate. In this arrangement, the thermal radiation of the heater has sufficient intensity near the band gap wavelength to serve as a radiation source for the transmission measurement even at temperatures below 200$^{\circ}$C, provided that no diffuser plate is inserted between the substrate and the heater. In Fig.~\ref{fig_rheed} we show examples of RHEED patterns for the nominally $x=7\%$ doped (Ga,Mn)As material grown at stoichiometric 1:1 ratio of As:(Ga+Mn) and for different substrate temperatures $T_s$. For this nominal doping and flux ratio, the transition between 3D and 2D growth modes occurs for $T_s$ between 240 and 220$^{\circ}$C. In Fig.~\ref{fig_3D-2D} we plot the 3D/2D growth boundary as a function of nominal Mn doping, reproducing essentially the previously reported observation \cite{Foxon:2004_a} that the maximum allowed $T_s$ for the 2D growth rapidly decreases with increasing doping. It can also be seen in Fig.~\ref{fig_3D-2D} that the 2D/3D transition temperature depends on the As:(Ga+Mn) ratio. The excess As flux compensates for the higher As re-evaporation rate and allows thus to increase the growth temperature without initiating the 3D growth mode. On the other hand, this gain in the maximum growth temperature, positive from the perspective of the epitaxial growth, does not fully compensate the increase in the density of As$_{\rm Ga}$ antisites, and the highest quality (Ga,Mn)As layers are still obtained when growing {(i)} in the 2D growth mode, {(ii)} as close as possible to the 2D/3D transition line, {(iii)} with the stoichiometric ratio As:(Ga+Mn)$\approx$1:1.

\vspace*{0cm}
\section{Film thickness and post-growth annealing}
\begin{figure}[!h]
\vspace*{0cm}
\hspace*{-0cm}\includegraphics[width=0.4\columnwidth,angle=0]{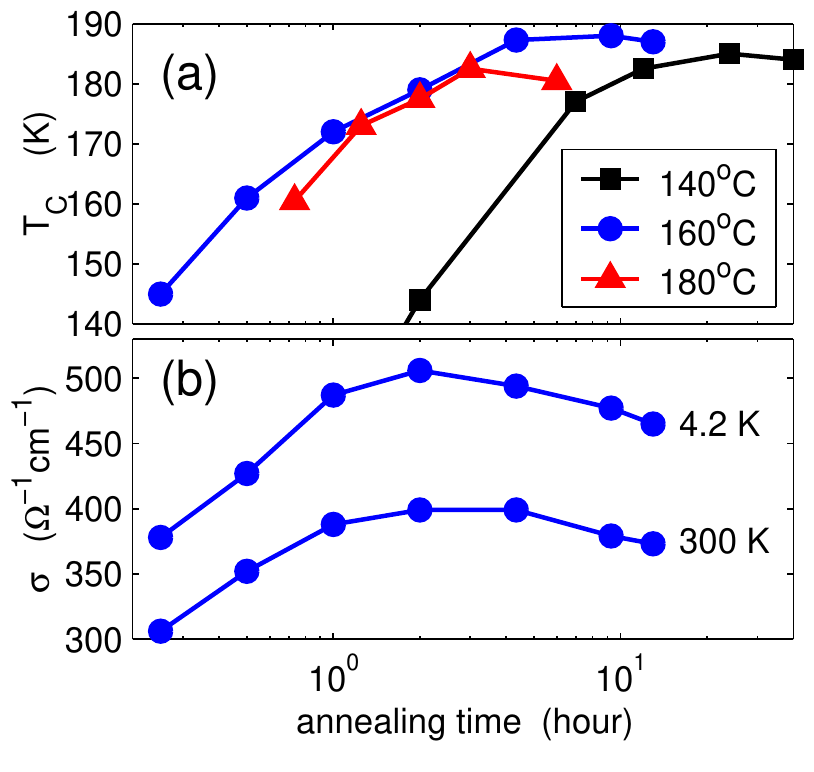}
\vspace*{0cm} \caption{(a) Curie temperature as a function of annealing time for three different annealing temperatures for a 15~nm thick (Ga,Mn)As epilayer with nominal 13\% Mn doping; the record $T_C=188$~K was achieved using annealing in air at 160$^\circ$C. (b) Conductivity (upper curve at 4.2~K and bottom curve at room temperature) monitored simultaneously during annealing at 160$^\circ$C of the same epilayer.}
\label{fig_anneal}
\end{figure}
Film thickness and post-growth annealing conditions are other important factors determining the quality of the resulting (Ga,Mn)As materials. In Fig.~\ref{fig_anneal} we show the dependence of the Curie temperature $T_c$ on the annealing time for three different annealing temperatures for our record $T_c=188$~K sample with nominal 13\% Mn doping and film thickness 15~nm. These curves illustrate the common trend in annealing (at temperatures close to the growth temperature) suggesting two competing mechanisms. One yielding the increase of $T_c$ and ascribed in a number of reports to the removal of charge and moment compensating interstitial Mn impurities (see also our detailed annealing studies in Ref.~\cite{Olejnik:2008_a}). In highly doped samples ($x\gtrsim 8$\%) a second mechanism is observed which starts to slowly reduce  $T_c$ after sufficiently long annealing times and at higher annealing temperatures. The origin of this detrimental mechanism is not established but can be ascribed to a partial removal of substitutional Mn or to a formation of some stable complexes (e.g. 2Mn$_{\rm Ga}$+Mn$_{\rm int}$ clusters) in the film. The lower annealing temperature the longer time is required for observing the downturn of $T_c$ but also longer time is required to reach the maximum $T_c$. Because of the two competing mechanisms, the absolutely highest Curie temperature for the given nominal doping is achieved at some intermediate annealing temperature and time, as illustrated in Fig.~\ref{fig_anneal}(a).
\begin{figure}[!h]
\vspace*{0cm}
\hspace*{-0cm}\includegraphics[width=0.4\columnwidth,angle=0]{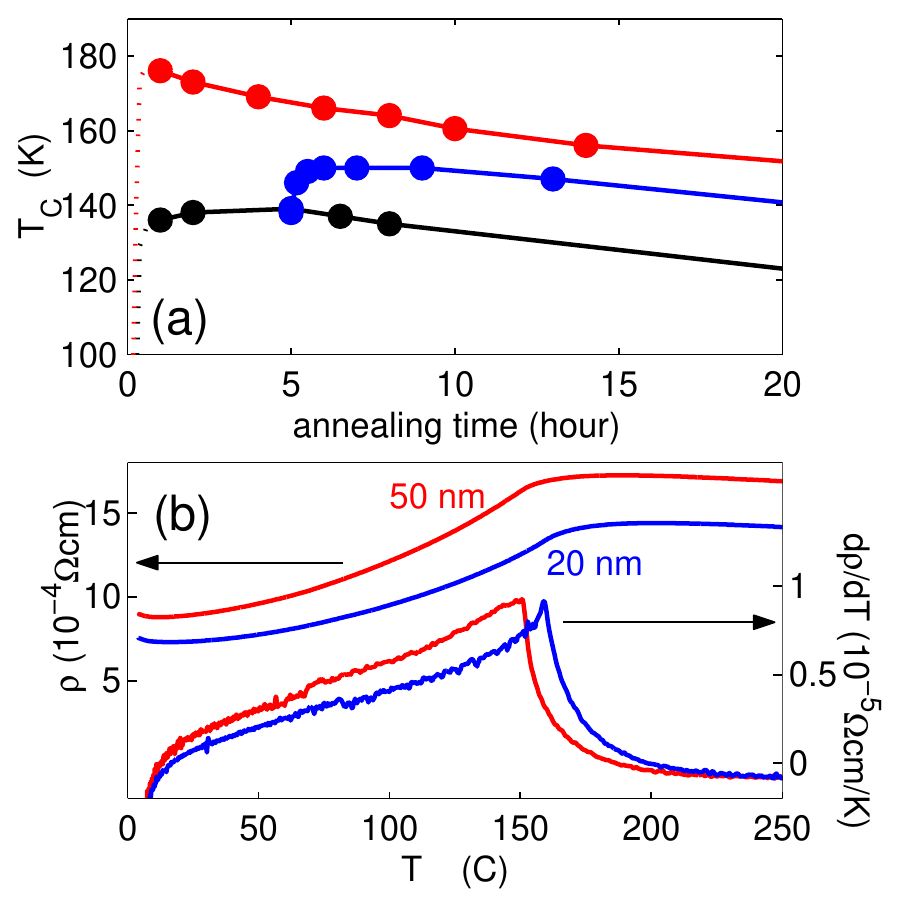}
\vspace*{0cm} \caption{(a) Annealing study of a 100~nm grown 13\% doped epilayer directly annealed (black curve), annealed after thinning (red curve) and partially annealed, thinned and re-annealed (blue curve). (b) Resistivities $\rho(T)$ and the $d\rho/dT$ singularities of 50 and 20~nm films with the same nominal Mn doping of 7\% showing good quality of optimally grown and annealed 50~nm films but not reaching fully the quality of optimized thinner films.}
\label{fig_thinning}
\end{figure}
\begin{figure}[!h]
\vspace*{0cm}
\hspace*{-0cm}\includegraphics[height=0.4\columnwidth,angle=0]{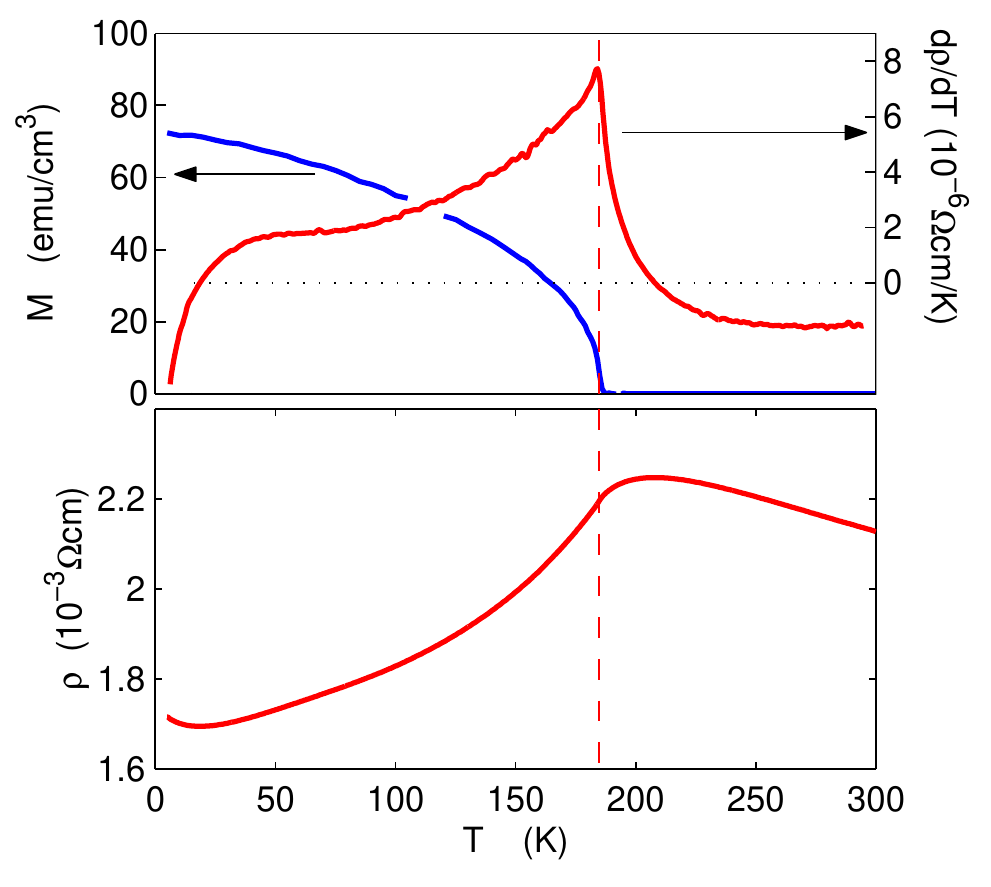}
\vspace*{0cm} \caption{Upper panel shows temperature dependent magnetization $M(T)$ and $d\rho/dT$  illustrating high magnetic quality of the optimized films; resistivity $\rho(T)$ is shown in the lower panel. Singularities in $M(T)$ and $d\rho/dT$ can be used to infer the Critical temperature while the broad maximum in $\rho(T)$ is above $T_c$. This is a common result illustrated in this figure on an optimized 11\% nominally doped (Ga,Mn)As film.}
\label{fig_drdt}
\end{figure}

We also note that while the increase of $T_c$ with annealing is closely correlated with the increase of the conductivity of the film, which is often used as a practical tool for monitoring the annealing progress, the maximum $T_c$ is typically reached at a later annealing time than the maximum of the conductivity, as illustrated in Fig.~\ref{fig_anneal}(b). This effect is stronger in very thin samples and can be ascribed to layer thinning due to oxidation during the annealing. The difference between the peak conductivity and the conductivity at maximum $T_c$ remain, however, small at practical annealing times.

For a given nominal doping, the highest attainable $T_c$ is reached only in thin films, typically thinner than $\sim$50~nm. (Note that in ultra-thin films with thicknesses below $\sim$10~nm, which start to show a clear enhancement of resistivity, $T_c$ also does not reach the highest values obtained at the given nominal doping in the thicker films.) In Fig.~\ref{fig_thinning}(a) we illustrate the importance of the film thickness for obtaining the high quality (Ga,Mn)As materials.  A 100~nm thick film is grown with nominal 13\% doping and, unlike the thin record $T_c$ film discussed above, here the maximum $T_c$ achieved by annealing is only about 140~K. However, even after passing this maximum the $T_c$ can still be increased when the annealed 100~nm thick film is thinned down (to, e.g., 20~nm) by wet etching and annealed further, as also shown in Fig.~\ref{fig_thinning}(a). The highest
$T_c$ is achieved when the film is thinned down before any annealing.

Our systematic studies of (Ga,Mn)As materials therefore focus on films with thickness below 50~nm and the most extensive series of carefully optimized materials was prepared using 20~nm thick epilayers. To illustrate that the 50~nm films can have good quality but do not fully reach the $T_c$, low resistivity, and sample uniformity after optimal annealing as the 20~nm films with the same nominal doping we show in Fig.~\ref{fig_thinning}(b) a comparison of resistivities, $\rho(T)$, and the $d\rho/dT$ singularities at the critical point for the 50 and 20~nm thick films (with 7\% nominal doping). The resistivity is smaller and the $d\rho/dT$ singularity is sharper in the 20~nm film confirming a higher degree of uniformity. The position of the $d\rho/dT$ singularity, corresponding to $T_c$ (see Fig.~\ref{fig_drdt} and Ref.~\cite{Novak:2008_a} for details), is shifted to slightly higher temperature in the thinner film.

\section{Series of ${\rm\bf (Ga,Mn)As}$ materials prepared under optimized growth and annealing conditions}
\begin{figure}[!h]
\vspace*{0cm}
\hspace*{-0cm}\includegraphics[width=0.4\columnwidth,angle=0]{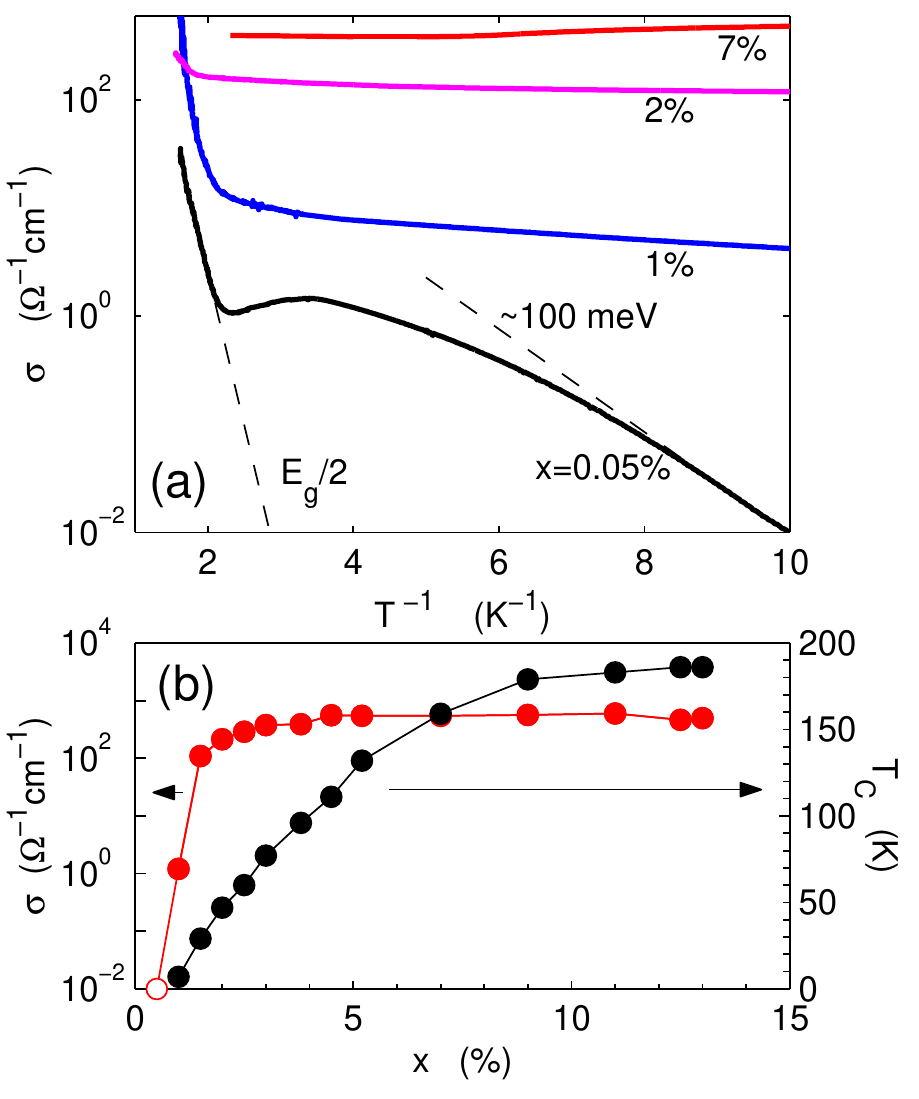}
\vspace*{0cm} \caption{Measurements of the insulator to metal transition induced by increasing Mn doping in (Ga,Mn)As. (a) 
Temperature dependent conductivities in the insulating low doped and metallic high doped samples.
(b) Red curve: conductivity at 4.2~K as a function of doping showing abrupt increase of the conductivity at $x\approx1.5\%$ and only a weak dependence (on this large scale) of conductivity on doping beyond this value. (b) Black curve: dependence of Curie temperature on doping.}
\label{fig_MIT}
\end{figure}

\begin{figure}[!h]
\vspace*{0cm}
\hspace*{-0cm}\includegraphics[width=0.4\columnwidth,angle=0]{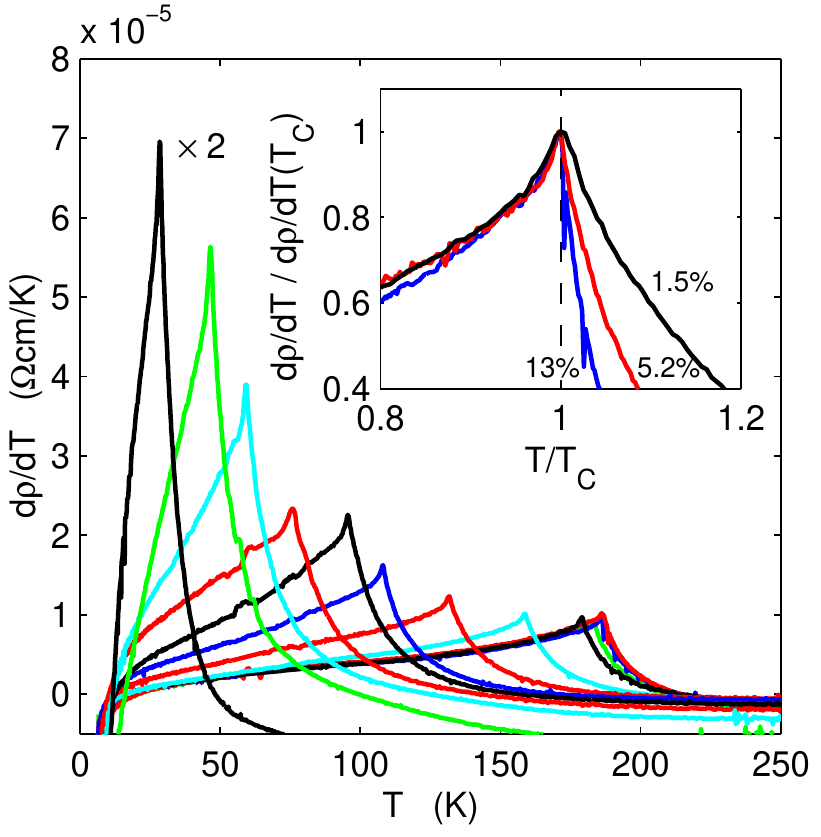}
\vspace*{0cm} \caption{$d\rho/dT$  measurements illustrating high magnetic quality of the optimized films throughout the whole series spanning the nominal doping range from $\sim$1.5 to 13\% and $T_c$'s from $\sim$25 to 186~K. In the inset detail of the $d\rho/dT$ cusp with maximum normalized to unity is shown.}
\label{fig_drdt_series}
\end{figure}
After finding the optimal growth and post-growth conditions for each individual nominal doping we obtained a series of samples whose characteristics are summarized in Tab.~\ref{tab_series}. The samples can be divided into several groups: at nominal dopings below $\sim0.1\%$ the (Ga,Mn)As materials are paramagnetic,  strongly insulating, showing signatures of the activated transport corresponding to valence band -- impurity band transitions at intermediate temperatures, and valence band -- conduction band transitions at high temperatures (see upper panel in Fig.~\ref{fig_MIT}). For higher nominal dopings, $0.5 \lesssim x \lesssim 1.5\%$, no clear signatures of activation from the valence band to the impurity band are seen in the dc transport, suggesting that the bands start to overlap and mix, yet the materials remain insulating. At $x\approx1.5\%$, low-temperature conductivity of the film increases abruptly by several orders of magnitude (see bottom panel of Fig.~\ref{fig_MIT}), and the system turns into a degenerate semiconductor. The onset of ferromagnetism occurs already on the insulating side of the transition at $x\approx 1\%$ and the Curie temperature then steadily increases with increasing doping up to nominal doping of $\approx13\%$. Beyond these dopings we did not succeed in introducing more substitutional Mn$_{\rm Ga}$ moments into the system.

In Fig.~\ref{fig_drdt_series} we show the $d\rho/dT$ measurements throughout the series of optimized samples spanning the nominal doping range from $\approx$1.5 to 13\% which illustrate the high quality of all the epilayers within the series.

\begin{figure}[!h]
\vspace*{0cm}
\hspace*{-0cm}\includegraphics[width=0.4\columnwidth,angle=0]{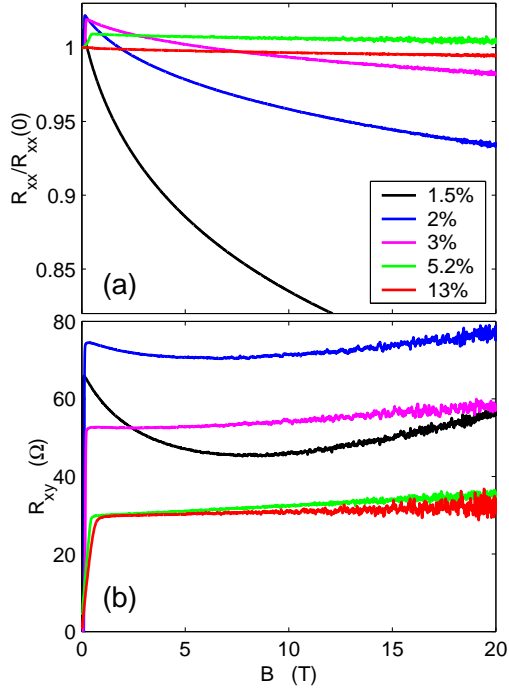}
\vspace*{0cm} \caption{Transport measurements of the series of optimized materials. (a) Longitudinal resistance of selected samples (normalized to $R_{xx}$ at $B = 0$). (b) Hall resistance; note the strong anomalous Hall-effect contribution.}
\label{fig_Hall1}
\end{figure}
\begin{figure}[!h]
\vspace*{0cm}
\hspace*{-0cm}\includegraphics[width=0.4\columnwidth,angle=0]{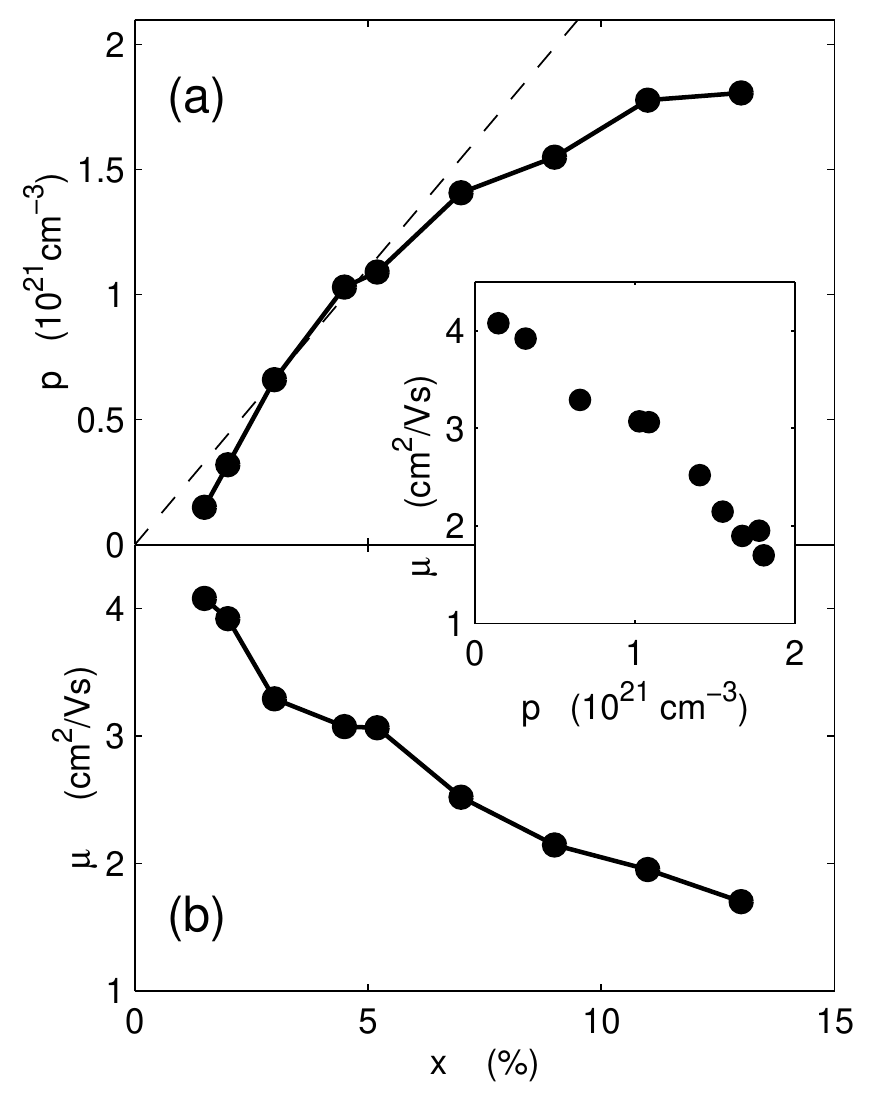}
\vspace*{0cm} \caption{(a) Hole density at 4.2~K extracted from the high-field slope of the Hall resistance. (b) Hole mobility at 4.2~K. Inset: Hole mobility vs. hole density.}
\label{fig_Hall2}
\end{figure}
\begin{figure}[!h]
\vspace*{.5cm}
\hspace*{-0cm}\includegraphics[width=.4\columnwidth,angle=0]{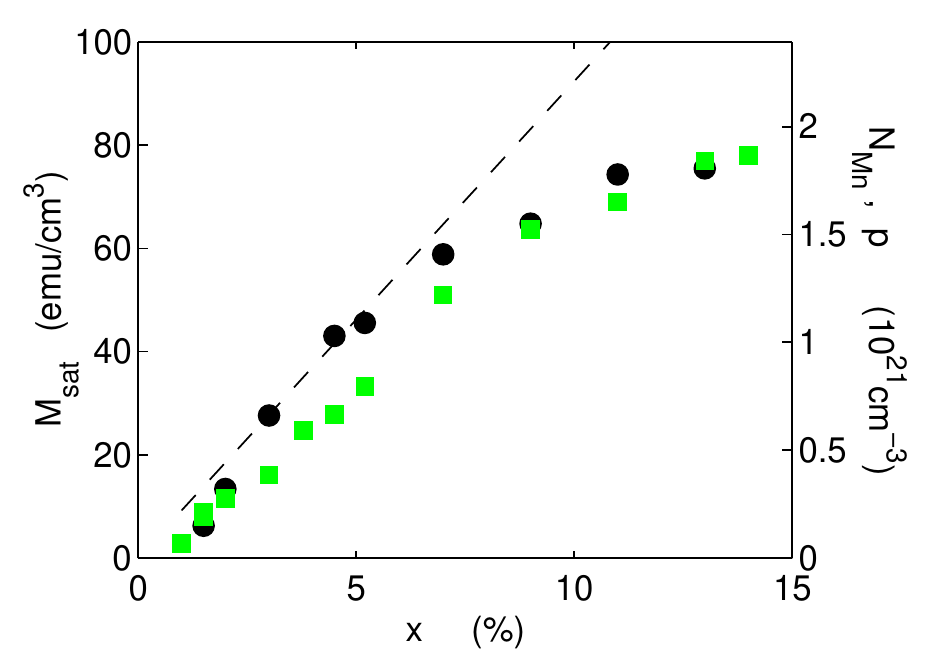}
\vspace*{0cm} \caption{Magnetization measurement of the series of optimized materials. Saturated moment is plotted by green squares and refers to the left y-axis.  Right y-axis translates the saturated moment into the density $N_{Mn}$ of uncompensated substitutional Mn atoms. Black circles denote the hole density $p$ inferred from the Hall effect. Dashed line indicates the case when the effective uncompensated doping equals to the nominal doping $x$, i.e. when $N_{Mn}=p=4x/a_L^3$.}
\label{fig_Msat}
\end{figure}

Fig.~\ref{fig_Hall1} shows results of the Hall effect measurements at 4.2~K. For this purpose the samples were lithographically patterned into Hall-bars of $60\mu$m width. It can be seen in Fig.~\ref{fig_Hall1} that the Hall signal is affected by longitudinal magnetoresistance of the samples, especially at low dopings. Therefore, we extracted $p$ from high field data and by fitting the measured transversal resistance
$R_{xy}$ by 
\begin{equation}
R_{xy}=B/(epd) + k_1R_{xx} + k_2R_{xx}^2
\end{equation}
where $d$ is the sample thickness and $k_1$ and $k_2$ are fitting constants
reflecting the anomalous Hall effect and possible imperfections in the geometry of the Hall bars. 

We also emphasize that, apart from the common experimental scatter and  from the corrections due to the non-zero magnetoresistance and due to the anomalous Hall effect, the carrier density can in principle be inferred only approximately from the slope of the Hall curve in a multi-band, spin-orbit coupled exchange-split system such as the (Ga,Mn)As.  The error bar due to the multi-band nature is estimated to $\sim 20\%$ \cite{Jungwirth:2005_b}. Due to  these uncertainties we can only make  semi-quantitative conclusions based on the measured Hall effect hole densities. Among those conclusions is the confirmed overall trend of increasing $p$ (and decreasing mobility $\mu$) with increasing doping in the optimized materials, as shown in Fig.~\ref{fig_Hall2}.  

Magnetization measurements are summarized in Fig.~\ref{fig_Msat}. The figure confirms the increasing trend in the saturation moment (corresponding to the density of uncompensated Mn$_{\rm Ga}$ moments) with increasing nominal doping up to $x\approx 13$. Assuming 4.5$\mu_B$ per Mn atom \cite{Jungwirth:2005_a} we can estimate the density $N_{Mn}$ of uncompesated Mn$_{\rm Ga}$ from the magnetization data, as shown by the right y-axis in Fig.~\ref{fig_Msat}. An important conclusion can be drawn when comparing this estimate with the hole density $p$ previously estimated from the Hall resistance. Since there is no apparent deficit of $p$ compared to $N_{Mn}$, and since the interstitial Mn impurity compensates one local moment but two holes we conclude that interstitial Mn is completely (within the experimental scatter) removed in our optimally annealed epilayers. The series of optimized materials is essentially free of charge and moment compensation and the density of both holes and local moments is given by the density of Mn in substitutional sites. As can be inferred from Fig.~\ref{fig_Msat}, the maximum effective doping of substitutional Mn atoms reaches $\approx 8$\% at nominal doping $x\approx 13$\%.


\begin{table*}[!h]
\vspace*{0cm}
\begin{tabular}{c | c c c c c c c}
sample  &
\,\, $x$ \,\, &
\,\, $d$ \,\, &
\,\, $T_C$ \,\, &
\,\, $\sigma$ at 300~K \,\, &
\,\, $\sigma$ at 4.2~K \,\, &
\,\, $p$ at 4.2~K \,\, &
\,\, $M_{sat}$ \\
     &
(\%) &
(nm) &
(K)  &
($\Omega^{-1}\rm{cm}^{-1}$) &
($\Omega^{-1}\rm{cm}^{-1}$) &
($10^{21}$cm$^{-3}$) &
(emu/cm$^3$) \\
\hline
E047 & 0.04 & 100 & 0 & 11 & 0 & 0 & 0\\
F031 & 0.1 & 100 & 0 & 37 & 0 & 0 & 0\\
F029 & 0.5 & 100 & 0 & 30 & 0 & 0 & 0\\
F033 & 1.0 & 100 & 7 & 82 & 1 & - & 2.8\\
F014 & 1.5 & 100 & 25 & 128 & 40 & - & 8.0\\
F010 & 1.5 & 20 & 29 & 156 & 98 & 0.15 & 8.9\\
F008 & 2.0 & 20 & 47 & 200 & 201 & 0.32 & 11.6\\
F007 & 2.5 & 20 & 60 & 233 & 291 & - & 11.5\\
F002 & 3.0 & 20 & 77 & 277 & 348 & 0.66 & 16.2\\
F016 & 3.8 & 20 & 96 & 286 & 401 & - & 24.7\\
E101 & 4.5 & 19 & 111 & 386 & 507 & 1.03 & 27.8\\
F020 & 5.2 & 20 & 132 & 395 & 535 & 1.08 & 33.3\\
E115 & 7.0 & 20 & 159 & 427 & 568 & 1.41 & 51.0\\
D071 & 7.0 & 50 & 150 & 400 & 552 & -    & 47.4 \\
E122 & 9.0 & 20 & 179 & 414 & 533 & 1.55 & 63.7\\
E094 & 11.0 & 20 & 183 & 464 & 556 & 1.78 & 69.0\\
E079 & 12.5 & 20 & 186 & 388 & 464 & - & 79.1\\
F055 & 13.0 & 20 & 186 & 405 & 492 & 1.81 & 76.9\\
E077 & 13.0 & 15 & 188 & 379 & 477 & - & -\\
F056 & 14.0 & 20 & 182 & 386 & 509 & - & 78.1\\
\hline
\end{tabular}
\vspace*{0cm} \caption{Table summarizing the basic characteristics of samples from the series of optimized materials.}
\label{tab_series}
\end{table*}

\vspace*{0cm}
\section{Optical experiments}
The infrared transmission spectra were recorded at 300~K using beam condenser kit on a Nicolet 6700 FTIR spectrometer with 4~cm$^{-1}$ resolution and Happ-Genzel apodization. The spectra in the MIR region (350-4000~cm$^{-1}$) were recorded using KBr beamsplitter and DTGS detector; the spectra in the NIR region (4000-11000~cm$^{-1}$) were recorded using CaF2 beamsplitter and PbSe detector. The spectra in the FIR region (25-700~cm$^{-1}$) were recorded on an evacuated Bruker IFS 113v spectrometer with 2~cm$^{-1}$ resolution and Blackman-Harris 3-Term apodization using 6 $\mu$m thick  mylar beamsplitter and a DTGS/PE detector. 

For the THz time-domain experiments we used the experimental setup described in Ref.~\cite{Kuzel:2007_a}. The linearly polarized broadband THz pulses formed a $\sim 3$~mm spot on the sample and probed the in-plane complex conductivity $\sigma$ of (Ga,Mn)As films. Each experiment consisted of two consecutive measurements: one of a time-domain signal $E_s(t)$ transmitted through the (Ga,Mn)As/GaAs wafer and another one of a reference signal $E_r(t)$ transmitted through a bare GaAs substrate. The Fourier transforms of the signals allow us to calculate the complex transmittance spectrum of the film: $t(\omega) = E_s(\omega)/E_r(\omega)$. The spectra of complex refractive index $N$ (and of the conductivity $\sigma =-i\omega\epsilon_0N^2$) are then retrieved following the procedure described in Ref.~\cite{Kuzel:2007_a}.
In our samples with the substrate polished from the back side, the wafer thickness precision is 1-2~$\mu$m. 
This leads to 
the experimental error of the order of 30-60 $\Omega^{-1}$cm$^{-1}$ in the magnitude of the measured Re~$\sigma$ in the THz range which is plotted as dots in Fig. 1(b) in the main text.

As emphasized in Ref.~\cite{Jungwirth:2007_a2}, no general practical sum rule exists for the low-frequency  part of the ac conductivity of (Ga,Mn)As films. Consequently, we did not attempt to interpret the shape of the observed low-frequency conductivity. We used a non-Drude (or generalized Drude) formula \cite{Beard:2000_a} which allowed us to obtain a very good fit of the data simultaneously in the THz and far-infrared frequency ranges. The conductivity peak in the mid-infrared region was fitted by adding  a sum of three damped harmonic oscillators (in order to take into account its asymmetry). The entire fitting formula then reads,
\begin{eqnarray}
\sigma(\omega)&=&\frac{\sigma_{dc}}{\left[1-(i\omega\tau)^{\alpha}\right]^{\beta}}\nonumber \\
& &-i\omega\epsilon_0\sum_{i=1}^3\frac{f_i}{\omega_i^2-\omega^2-i\omega\gamma_i}-i\omega\epsilon_0\epsilon_{\infty}
\end{eqnarray}
We obtained excellent simultaneous fits of the THz, far- and mid-infrared transmittance data by using the above formula (with $\alpha\neq1$ and $\beta\neq1$) and the transfer matrix formalism \cite{Jacobsson:1965_a} to calculate the transmission through (Ga,Mn)As/GaAs system. We emphasize that the uncertainty in the resulting mid-infrared peak position of Re~$\sigma$ is much smaller than the shift of the peak observed across the studied Mn doping range.

The spectra of MCD were studied by the experimental procedure described in \cite{Sato:1981_a}. In particular, the photoelastic modulator (PEM) was used to alter the polarization state of the light. The sign of MCD was determined from an independent experiment, where the helicity of the light was controlled by a quarter wave plate. All MCD experiments were performed at temperature of  15~K. The angle of incidence of the light beam was smaller than 1~deg (i.e., very close to the normal incidence). An external magnetic field of 540~mT was applied perpendicular to the sample plane during the experiment. The magnetic field direction-independent effects were removed by taking the difference between the data measured with +540~mT  and -540~mT. The data were also corrected for the MCD signal from the cryostat windows.

\section{Parabolic model of optical dielectric tensor in metallic ${\rm\bf GaMnAs}$}

The interband conductivity of a system can be calculated through the Kubo formula :
\begin{eqnarray}
\sigma_{\alpha\beta}^{\rm inter}(\omega)
&=&\frac{i e^2}{m_0^2 \omega V}
\sum_{\kk,a\ne b}(f_{a,k}-f_{b,k})\frac{p_{ab}^\alpha p_{ba}^\beta}
{\hbar \omega-E_{ba}+i\eta}
\end{eqnarray}
where $p^\alpha_{ab}=\langle a k|\hat{p}_\alpha|bk\rangle$ and $E_{ba}=E_{bk}-E_{ak}$. Its relation to the dielectric tensor is
\begin{equation}
\epsilon_{\alpha,\beta}=\epsilon_\infty\delta_{\alpha,\beta}+\frac{i 4\pi}{\omega}\sigma_{\alpha,\beta}
\end{equation}
which obey the Kramers-Kronig relations
\begin{eqnarray}
{\rm Re}[\epsilon_{\alpha\beta}-\epsilon_\infty\delta_{\alpha\beta}]&=&
\frac{1}{\pi}P\int_{-\infty}^\infty \frac{{\rm Im}[\epsilon_{\alpha\beta}]}{\omega'-\omega}d\omega'=
\frac{2}{\pi}P\int_0^\infty \frac{\omega'{\rm Im}[\epsilon_{\alpha\beta}]}{{\omega'}^2-\omega^2}d\omega'\\
{\rm Im}[\epsilon_{\alpha\beta}]&=&
-\frac{1}{\pi}P\int_{-\infty}^\infty \frac{{\rm Re}[\epsilon_{\alpha\beta}]}{\omega'-\omega}d\omega'=
-\frac{2}{\pi}P\int_0^\infty \frac{\omega{\rm Re}[\epsilon_{\alpha\beta}]}{{\omega'}^2-\omega^2}d\omega'
\end{eqnarray}

Ignoring the dc-contributions to the dielectric constant, the dielectric tensor from inter-band contributions can be written as
\begin{eqnarray}
\epsilon_{\alpha\beta}^{\rm inter}(\omega)-\epsilon_\infty\delta_{\alpha,\beta}
&=&
\frac{4\pi e^2\hbar^2}{m^2 V}
\sum_{\kk,a\ne b}\frac{(f_{a,k}-f_{b,k})}{E_{ba}^2}\frac{p_{ab}^\alpha p_{ba}^\beta}
{E_{ba}-\hbar \omega-i\eta}
\end{eqnarray}
This expressions amounts to saying that one requires $\epsilon_{xy}^{\rm inter}(\omega=0)=0$
and that $\epsilon_{xx}^{\rm inter}(\omega=0)=\frac{4\pi e^2\hbar^2}{m^2 V}
\sum_{\kk,a\ne b}\frac{(f_{a,k}-f_{b,k})p_{ab}^x p_{ba}^x}{E_{ba}^3}$.
In calculating the dielectric tensor it is simpler to calculate first the contributions containing delta functions and then obtain the other contributions through the KK relations. 
More specifically
\begin{eqnarray}
{\rm Im}[\epsilon_{xx}^{\rm inter}(\omega)]
&=&\frac{4\pi^2 e^2\hbar^2}{m^2  V}\sum_{\kk,a\ne b}
\frac{(f_{a,k}-f_{b,k})}{E_{ba}^2}
{\rm Re}[p_{ab}^x p_{ba}^x]
\delta(\hbar \omega-E_{ba})\
\label{Resigma_xx}
\end{eqnarray}
and
\begin{eqnarray}
{\rm Re}[\epsilon_{xy}^{\rm inter}(\omega)]
&=&-\frac{4\pi^2 e^2\hbar^2}{m^2  V}\sum_{\kk,a\ne b}
\frac{(f_{a,k}-f_{b,k})}{E_{ba}^2}
{\rm Im}[p_{ab}^\alpha p_{ba}^\beta]
\delta(\hbar \omega-E_{ba})
\label{Imsigma_xy}
\end{eqnarray}

\begin{figure}[h]
\includegraphics[width=0.5\columnwidth]{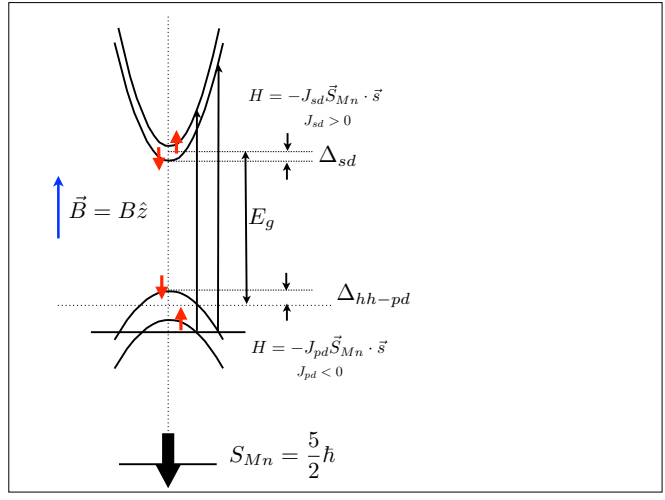}
\caption{ Sketch of parabolic model (only two valence bands shown).}
\label{sketch}
\end{figure}

{\it Kinetic-exchange parabolic band model.} The model that we consider is one of parabolic bands with effective masses $m_c$,$m_{hh}$,$m_{lh}$,$m_{so}$. The conduction bands are unoccupied and the valence bands are partially occupied with a given Fermi energy $E_F$ measured from the top of the valence band without exchange splitting as shown in Fig.~\ref{sketch}. 
Here the label $c \in b$ and $hh,lh,so \in a$. The exchange splitting (see Fig.~\ref{sketch}) is given by $\Delta_{ba}=\Delta_{sd}s_b+\Delta_{pd}s_a$. Here $\Delta_{sd}$ is due to the ferromagnetic direct-exchange coupling between the conduction band $s$-orbitals and the Mn $d$-orbitals from the same Ga-sublattice of the host zinc-blende crystal; $\Delta_{pd}$ is due to the antiferromagnetic kinetic-exchange coupling which originates from the hybridization between the $p$-orbitals from the As sublattice and the Mn $d$-orbitals.  $\Delta_{sd}$ and $\Delta_{pd}$ are defined as positive numbers.

In what follows we consider pairs of valence bands and conduction bands (i.e. the transitions of the $\pm 3/2 hh$ to the $\pm 1/2 c$, etc.) with the labeling:
\begin{eqnarray}
E_{b}&=&\frac{\hbar^2k^2}{2m_c}+\Delta_{sd}s_b+E_g\\
E_{a}&=&-\frac{\hbar^2k^2}{2m_a} -\Delta_{pd}s_a\\
E_{ba}&=&E_a-E_b=\frac{\hbar^2k^2}{2m_c}+\frac{\hbar^2k^2}{2m_a} +\Delta_{ba}=
\frac{\hbar^2k^2}{2\mu_ab}+\Delta_{ba}
\end{eqnarray}
where $\mu_{ab}=m_a m_c/(m_a+m_c)$. We will also use the usual notation for the dipole matrix elements $E_P\equiv 2m_0 P^2/\hbar^2$
and $P=-(i\hbar/m_0)\langle S|\hat{p}_x|X\rangle$.

Then for the Imaginary component of the diagonal dielectric tensor the contribution from the transition from $a$  to $b$ ($f_b=0$) is:
\begin{eqnarray}
{\rm Im}[\epsilon_{xx}^{\rm a-b}(\omega)]^{\rm clean}
&=&\frac{2E_P\sqrt{\rm Ry}}{  (\hbar\omega)^2  }\eta_{ab}^{xx}\left(\frac{\mu_{ab}}{m_0}\right)^{3/2}
\sqrt{(\hbar \omega-\Delta_{ba})} \theta(\hbar\omega-\frac{\hbar^2 k_{Fa}}{2\mu_{ab}}-\Delta_{ba})\\
&\equiv& \frac{f_{ab}}{(\hbar\omega)^2} \sqrt{(\hbar \omega-\Delta_{ba})} \theta(\hbar\omega-\frac{\hbar^2 k_{Fa}}{2\mu_{ab}}-\Delta_{ba})
\end{eqnarray}
where we have used ${\rm Ry}=\hbar^2/2m_0a_0^2=e^2/2a_0$ and $e^2/4ha_0=e^2/8\pi\hbar a_0=1.83\times 10^3 \Omega^{-1}{\rm cm}^{-1}$.

The real part of $\epsilon_{xx}$ can then be found from the KK relations:
\begin{eqnarray}
{\rm Re}[\epsilon_{xx}-\epsilon_\infty]^{\rm clean}&=&
\frac{2}{\pi}P\int_0^\infty \frac{\omega'f_{ab} \sqrt{(\hbar \omega'-\Delta_{ba})} \theta(\hbar\omega'-\frac{\hbar^2 k_{Fa}^2}{2\mu_{ab}}-\Delta_{ba})}{(\hbar\omega')^2({\omega'}^2-\omega^2)}d\omega' 
=\frac{2f_{ab}\sqrt{\Delta_{ba}} }{\pi\Delta_{ba}^2} P\int_{\frac{E_{ab-F}}{\Delta_{ba}}}^\infty 
\frac{ \sqrt{x-1}}{x(x^2-(\hbar\omega/\Delta_{ba})^2)}dx\nonumber\\
&=&\frac{4E_P\eta_{ab}^{xx}}{\pi}\left(\frac{\mu_{ab}}{m_0}\right)^{3/2}\frac{\sqrt{\Delta_{ba}{\rm Ry}}}{(\hbar\omega)^2}
\left[\pi-2\arctan\left[\sqrt{\frac{\epsilon_{ab-F}}{\Delta_{ba}}}\right]
\right.\nonumber\\&&\left.
-\sqrt{1-\frac{\hbar\omega}{\Delta_{ba}}}\left(\frac{\pi}{2}\theta(1-\frac{\hbar\omega}{\Delta_{ba}})-\arctan\left[\sqrt{\frac{\epsilon_{ab-F}}{{\Delta_{ba}}-\hbar \omega}}\right]\right)
\right.\nonumber\\&&
\left.-\sqrt{1+\frac{\hbar\omega}{\Delta_{ba}}}\left(\frac{\pi}{2}-\arctan\left[\sqrt{\frac{\epsilon_{ab-F}}{{\Delta_{ba}}+\hbar \omega}}\right]\right)
\right]
\end{eqnarray}
where $\eta\equiv \hbar\omega/\Delta_{ba}$, $E_{ab-F}=\frac{\hbar^2 k_{Fa}^2}{2\mu_{ab}}+\Delta_{ba}$ and $\epsilon_{ab-F}=\frac{\hbar^2k_{Fa}^2}{2\mu_{ab}}$.
\begin{eqnarray}
{\rm Re}[\epsilon_{xy}^{ a-b}(\omega)]^{\rm clean}
&=&-\frac{2E_P\sqrt{\rm Ry}}{  (\hbar\omega)^2  }\eta_{ab}^{xy}\left(\frac{\mu_{ab}}{m_0}\right)^{3/2}
\sqrt{(\hbar \omega-\Delta_{ba})} \theta(\hbar\omega-\epsilon_{ab-F}-\Delta_{ba})\\
&\equiv& -\frac{f^{xy}_{ab}}{(\hbar\omega)^2} \sqrt{(\hbar \omega-\Delta_{ba})} \theta(\hbar\omega-\epsilon_{ab-F}-\Delta_{ba})
\end{eqnarray}
\begin{eqnarray}
{\rm Im}[\epsilon_{xy}^{a-b}(\omega)]_{\rm }^{\rm clean}
&=&\frac{  2}{\pi } 
\frac{f_{ab}^{xy}}{(\hbar\omega)^2}\left[ 
-\frac{\hbar\omega\sqrt{\epsilon_{ab-F}}}{\epsilon_{ab-F}+\Delta_{ba}}
-\frac{\hbar\omega}{\sqrt{\Delta_{ba}}}
\left(\frac{\pi}{2}-\arctan\left[\sqrt{\frac{\epsilon_{ab-F}}{\Delta_{ba}}}\right]\right)
\right.\\&&+\left.
\sqrt{\Delta_{ba}+\hbar\omega}\left(\frac{\pi}{2}-\arctan\left[\sqrt{\frac{\epsilon_{ab-F}}{\Delta_{ba}+\hbar\omega}}\right]\right)
-\sqrt{\Delta_{ba}-\hbar\omega}\left(\frac{\pi}{2}
-\arctan\left[\sqrt{\frac{\epsilon_{ab-F}}{\Delta_{ba}-\hbar\omega}}\right]\right)
\right]\nonumber
\end{eqnarray}

\begin{center}
  \begin{tabular}{| c | c | c|c |c| c|}
    \hline
    &&&&&\\$a$ & $b$ & Re[$p_{ab}^x p_{ab}^x]\equiv M_{ab}^{xx}$ &$\eta^{xx}_{ab}\equiv \frac{2M_{ab}^{xx}}{m_0 E_P}$&Im[$p_{ab}^x p_{ab}^y]\equiv M_{ab}^{xy}$&$\Delta_{ba}$\\&&&&&\\ \hline\hline
    &&&&&\\hh +3/2 & c+1/2 & $\frac{m_0^2P^2}{2\hbar^2}=\frac{1}{2}\left(\frac{m_0E_P}{2}\right)$&$\frac{1}{2}$&$\frac{m_0^2P^2}{2\hbar^2} $&$\frac{1}{2}\Delta_{sd}+\frac{3}{2}\Delta_{pd}+E_g$\\&&&&&\\ \hline
    &&&&&\\hh -3/2 & c-1/2 & $\frac{m_0^2P^2}{2\hbar^2}=\frac{1}{2}\left(\frac{m_0E_P}{2}\right)$&$\frac{1}{2}$&-$\frac{m_0^2P^2}{2\hbar^2}$&$-\frac{1}{2}\Delta_{sd}-\frac{3}{2}\Delta_{pd}+E_g$\\&&&&& \\ \hline
    &&&&&\\lh +1/2 & c-1/2 &$ \frac{m_0^2P^2}{6\hbar^2}=\frac{1}{6}\left(\frac{m_0E_P}{2}\right)$&$\frac{1}{6}$&$\frac{m_0^2P^2}{6\hbar^2} $&$-\frac{1}{2}\Delta_{sd}+\frac{1}{2}\Delta_{pd}+E_g$\\&&&&&\\ \hline
    &&&&&\\lh -1/2 & c+1/2 &$\frac{m_0^2P^2}{6\hbar^2}=\frac{1}{6}\left(\frac{m_0E_P}{2}\right)$&$\frac{1}{6}$&-$\frac{m_0^2P^2}{6\hbar^2}$ &$\frac{1}{2}\Delta_{sd}-\frac{1}{2}\Delta_{pd}+E_g$\\&&&&&\\ \hline
   &&&&&\\ so +1/2 & c-1/2 & $\frac{m_0^2P^2}{3\hbar^2}=\frac{1}{3}\left(\frac{m_0E_P}{2}\right)$&$\frac{1}{3}$&$\frac{m_0^2P^2}{3\hbar^2} $&$-\frac{1}{2}\Delta_{sd}+\frac{1}{2}\Delta_{pd}+\Delta_{so}+E_g$\\&&&&&\\ \hline
    &&&&&\\so -1/2 & c+1/2 &$ \frac{m_0^2P^2}{3\hbar^2}=\frac{1}{3}\left(\frac{m_0E_P}{2}\right)$&$\frac{1}{3}$&-$\frac{m_0^2P^2}{3\hbar^2}$&$\frac{1}{2}\Delta_{sd}-\frac{1}{2}\Delta_{pd} +\Delta_{so}+E_g$\\&&&&&\\ \hline
  \end{tabular}
\end{center}

{\em Effects of disorder broadening.} Ferromagnetic (Ga,Mn)As is a very highly doped and therefore strongly disordered semiconductor material. This affects the magneto-optical effects by introducing a finite life time broadening  to the quasiparticles  and by allowing momentum non-conserving transitions. The former can be easily approximated by introducing an imaginary part to the energy denominators,  i.e., $i\delta\rightarrow i\Gamma$ (we used 100~meV). On the other hand, within our approximation, the effects of non-conserving momentum transitions translate in a lower bound of the transitions which we approximate by increasing the conduction band effective mass, $m_c\sim 0.7 m_0$. Also, the band gap renormalization from the electron-electron interactions, considering the larger effective mass, is given by $54\times (p[10^{18} {\rm cm}^{-3}])^{1/3}$ meV \cite{Sernelius:1986_a}. This leads to the above expressions being replaced by

\begin{eqnarray}
{\rm Re}[\epsilon_{xx}^{\rm inter}(\omega)]_{\rm dis}
&=&\epsilon_\infty +\frac{1}{\pi}\int_{-\infty}^{\infty} d(\hbar \omega'){\rm  Im}[\epsilon_{xx}^{\rm inter}(\hbar \omega')]^{\rm clean}
\frac{(\hbar\omega'-\hbar\omega)}{(\hbar\omega'-\hbar\omega)^2+\Gamma^2}
\end{eqnarray}

\begin{eqnarray}
{\rm Im}[\epsilon_{xx}^{\rm inter}(\omega)]_{\rm dis}
&=&\frac{1}{\pi}\int_{-\infty}^{\infty} d(\hbar \omega'){\rm Im}[\epsilon_{xx}^{\rm inter}(\hbar \omega')]^{\rm clean}
\frac{\Gamma}{(\hbar\omega'-\hbar\omega)^2+\Gamma^2}
\end{eqnarray}

\begin{eqnarray}
{\rm Re}[\epsilon_{xy}^{\rm inter}(\omega)]_{\rm dis}
&=&\frac{1}{\pi}\int_{-\infty}^{\infty} d(\hbar \omega'){\rm Re}[\epsilon_{xy}^{\rm inter}(\hbar \omega')]^{\rm clean}
\frac{\Gamma}{(\hbar\omega'-\hbar\omega)^2+\Gamma^2}
\end{eqnarray}

\begin{eqnarray}
{\rm Im}[\epsilon_{xy}^{\rm inter}(\omega)]_{\rm dis}
&=&-\frac{1}{\pi}\int_{-\infty}^{\infty} d(\hbar \omega'){\rm  Re}[\epsilon_{xy}^{\rm inter}(\hbar \omega')]^{\rm clean}
\frac{(\hbar\omega'-\hbar\omega)}{(\hbar\omega'-\hbar\omega)^2+\Gamma^2}
\end{eqnarray}



\end{document}